\documentclass[a4paper,12pt]{article}

\usepackage{graphics}

\newcommand{\pict}[2]
{
\begin{figure}[htb]
    \resizebox{\textwidth}{!}{
        \includegraphics{#1}
        }
    \caption{#2\label{fg:#1}}
\end{figure}
}

\begin{document}
\title{The (absence of a) relationship between thermodynamic and logical reversibility}
\author{O J E Maroney\\
Imperial College London\\
Physics Department\\
The Blackett Laboratory\\
Prince Consort Road\\
London\\
SW7 2BW} \maketitle

\begin{abstract}
Landauer erasure seems to provide a powerful link between
thermodynamics and information processing (logical computation).
The only logical operations that require a generation of heat are
logically irreversible ones, with the minimum heat generation
being $kT \ln 2$ per bit of information lost. Nevertheless, it
will be shown logical reversibility neither implies, nor is
implied by thermodynamic reversibility.  By examining
thermodynamically reversible operations which are logically
irreversible, it is possible to show that
 information and entropy, while having the same form, are
conceptually different.
\end{abstract}

\section{Introduction}\label{introduction}

In a recent article on the information theoretic exorcism of
Maxwell's demon, Bennett writes:

\begin{quote}
Landauer's principle, often regarded as the basic principle of the
thermodynamics of information processing, holds that any logically
irreversible manipulation of data \ldots must be accompanied by a
corresponding entropy increase in non-information bearing degrees
of freedom of the information processing apparatus or its
environment. Conversely, it is generally accepted any logically
reversible transformation of information can in principle be
accomplished by an appropriate physical mechanism operating in a
thermodynamically reversible fashion.(Bennett, 2003)
\end{quote}

In Bennett's, and in other papers (see Leff and Rex, 1990,2003;
Zurek 1990) it appears that a strong claim is being made linking
thermodynamic entropy to information processing in computers.  In
particular, it has been argued that the resolution of the Szilard
Engine(Szilard, 1929) requires an understanding of principles of
information processing.

In this paper it will be argued that, while Landauer erasure
(Landauer, 1961) is essentially correct (contrary, for example, to
the arguments of Earman and Norton (1998, 1999)) it has been
generalised incorrectly in its popularisation as "Landauer's
Principle". When properly understood, it will be shown that
information processing is unnecessary for resolving Maxwell's
demon and that the strong connection between information and
thermodynamic entropy is broken.

Two widespread generalisations of Landauer's paper, that may be
concluded from Bennett's quote above, can be stated as:
\begin{itemize}
\item All logically irreversible operations require a generation
of heat of $kT\ln 2$ energy per bit of information lost. \item
While all logically reversible operations may be accomplished as a
thermodynamically reversible process, some logically irreversible
operations cannot.
\end{itemize}
We will be argue that neither generalisation is correct:
\begin{itemize}
\item Some logically irreversible operations can be implemented
without generating heat. \item Any logically irreversible
transformation of information can in principle be accomplished by
an appropriate physical mechanism operating in a thermodynamically
reversible fashion.
\end{itemize}

In Section \ref{computation} the implications of this for the
thermodynamics of computation are considered, and then in Sections
\ref{shannon} and \ref{szilard} this is applied to Shannon
signalling and the Szilard Engine, respectively.  The conclusion
is that, while Shannon information and Gibbs entropy share the
same measure, they are not equivalent, and that principles of
information processing are not necessary to `save' the second law
of thermodynamics.
\section{Landauer Erasure}\label{landauer}

Landauer Erasure ($LE$) is a process by which a single information
bearing degree of freedom, in one of two possible logical states,
is restored to the logical value zero regardless of which state it
initially occupies (Figure \ref{fg:figure1}).

\pict{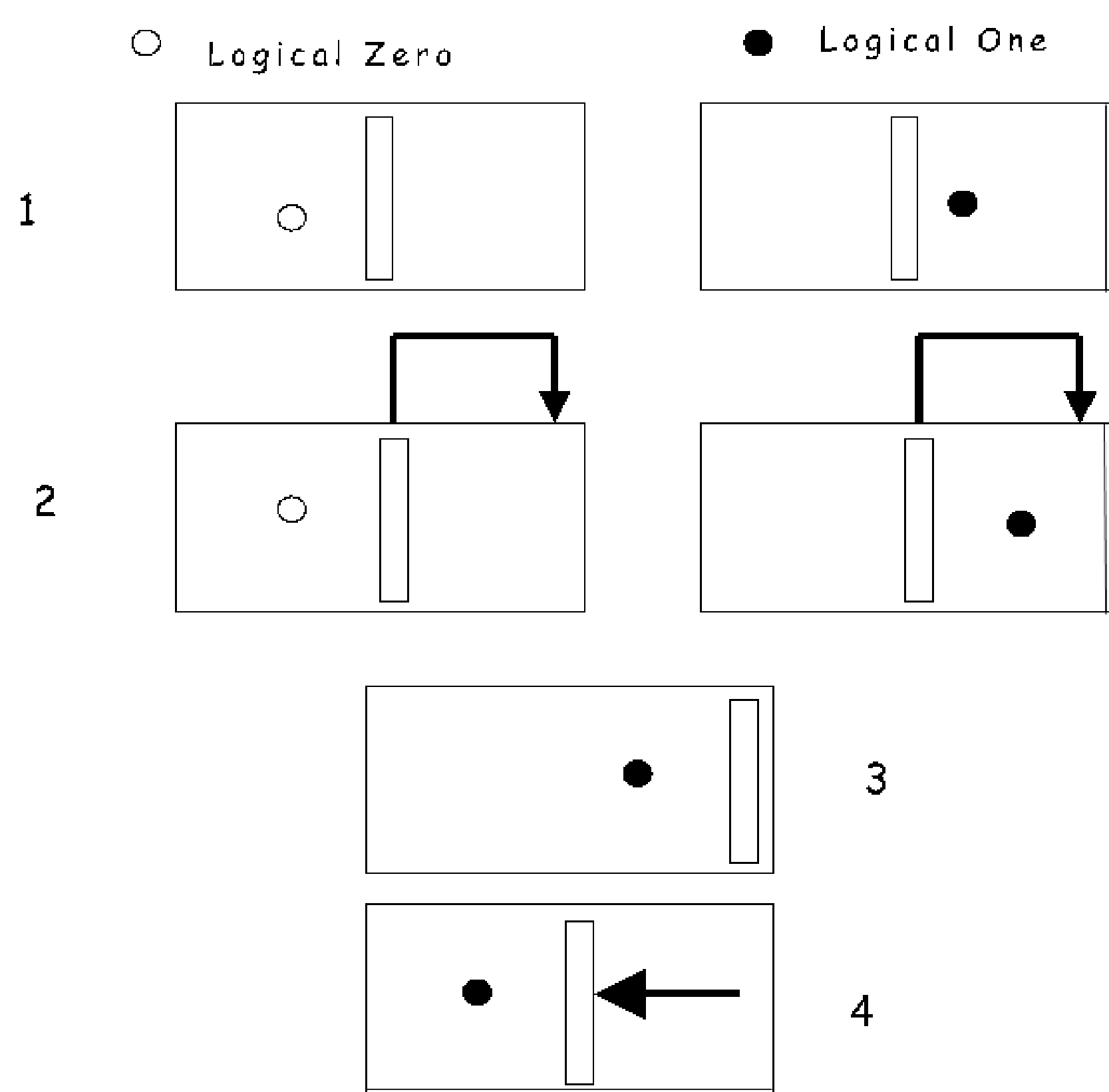}{Landauer Erasure}

\begin{enumerate}
\item The logical state may be represented by an `atom in a box',
with the box divided in two by a moveable barrier.  The atom on
the left represents logical state zero and on the right, logical
state one.  The box is in contact with a heat bath at some
temperature $T$. \item The barrier is removed from the box and the
atom is allowed to thermalise. \item The barrier is inserted in
the far right hand end of the box. \item The barrier is
isothermally moved to the centre of the box.  This acts as the
compression of an ideal gas so the work required is $NkT \ln 2$,
where $N=1$.  The work done on compressing the one atom gas is
expelled into the heat bath\footnote{For justification that the
ideal gas laws are valid for a single atom gas, see Leff (1995);
Bender, Brody and Meister (2000) and Maroney (2002)}.
\end{enumerate}

There are two features of this process that should immediately be
recognised:
\begin{itemize}
\item The compression stage (which requires the generation of
heat) is not the same as the erasure stage (which actually
destroys the information). It is noticeable that in their
exposition of $LE$, Leff and Rex (1990 p.22-25) refer to
"erasure/resetting". Stage 2 is the erasure, while Stage 4 is the
resetting.  A conclusion of our analysis will be that "erasure"
must be understood as a separate operation to "resetting". \item
The compression stage, Step 4 of the process, is an isothermal
compression, and is thermodynamically reversible.  The heat
generation is not a thermodynamically irreversible process.
\end{itemize}

\subsection{Reversing Landauer Erasure}
Let us now consider the effect of reversing these Stages, in a
"Reverse Landauer Erasure" ($RLE$).
\begin{enumerate}
\item The barrier is moved isothermally, reversibly to the far
right of the box.  Pressure from the expansion of the one atom gas
can be used to extract $kT \ln 2$ energy from the heat bath and
used as work. \item The barrier is removed from the right hand end
of the box. \item After a thermalisation period, the barrier is
inserted in the centre of the box. \item The atom has an equal
probability of being found on either side of the barrier.
\end{enumerate}
It is apparent that if the logical states of the atom were
initially uniformly distributed (ie. had probability $\frac{1}{2}$
of being on either side of the barrier) then the statistical state
of the system is now exactly the same as it was before $LE$.  The
work used to reset the bit in Stage 4 of $LE$ is recovered in
Stage 1 of $RLE$ and the entire process has taken place in a
thermodynamically reversible manner.

While this point might be accepted by advocates of the
information-entropy link, it is sometimes claimed that this is
only true if the initial state of the system has uniform
probability distribution:
\begin{quote}
memory erasure and resetting is always logically irreversible but
it is thermodynamically reversible only when the initial memory
ensemble is distributed uniformly among [0] and [1] states (Leff
and Rex 1990)
\end{quote}
If the initial statistical state of the bit is non-uniform, the
RLE process will not leave the final statistical state in the
initial distribution - the final state will still be probability
half of being in either statistical state.  As any non-uniform
distribution will have lower entropy, the result is an
uncompensated increase in Gibbs entropy and so is
thermodynamically irreversible.

\subsection{Non-uniform probabilities}
We will now show that the resetting of non-uniform probability
distributions can also take place as a thermodynamically
reversible process.

The key element here is to notice that the amount of information
one wishes to reset is less for a non-uniform probability:
\[
H=-\sum p \log_2 p
\]
This is 1 bit if the probabilities are equal, but less than 1 bit
if the probabilities are non-uniform.  As Landauer's principle
states only that $k T \ln 2$ energy must be dissipated {\em per
bit} of information, we should expect that with non-uniform
probabilities the erasure and resetting may take place with a
lower generation of heat.

\pict{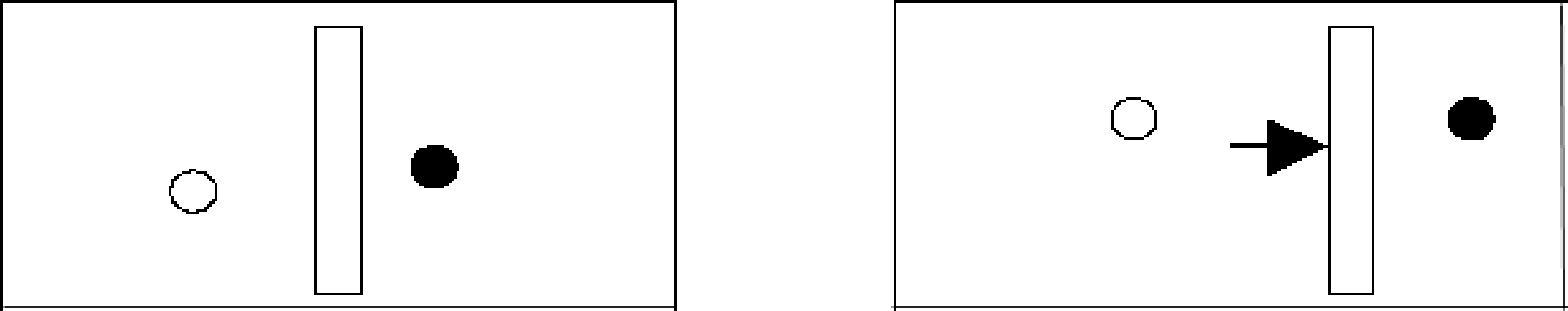}{Landauer Erasure of Non-Uniform Probabilities}

To achieve this lower generation it is necessary only to insert an
additional Stage (Figure \ref{fg:figure2}) between Stages 1 and 2
of the normal $LE$:
\begin{itemize}
\item Isothermally, reversibly move the barrier to the position
$x=p$ (where $x=0$ represents the barrier on the far left and
$x=1$ the far right of the box).
\end{itemize}
If the atom is initially located to the left of the piston, with
probability $p$, this requires a work expenditure of
\[
E_L=-kT \ln \left[ 2p \right]
\]
while if it is initially to the right, the work expenditure is
\[
E_R=-kT \ln \left[2 \left[1-p\right] \right]
\]
The mean expenditure of energy (over an ensemble of such
operations) is
\[
E=pE_L+(1-p)E_R=(H-1) k T \ln 2
\]
When combined with the work in Stage 4 of $LE$, the total work
required is
\[
W=H k T \ln 2
\]
in complete agreement with Landauer's principle \footnote{See
Shizume (1995), Piechocinska (2000) for proof that this is the
least work required.}

Let us call this process $LE(p)$ to emphasise that the erasure and
resetting operation is dependant upon the {\em amount} of
information one wishes to erase.

If we now reverse $LE(p)$ we find that the operation $RLE(p)$
modifies $RLE$ as follows:
\begin{itemize}
\item When the barrier is re-inserted (Stage 3) it is inserted at
the location $x=p$ in the box.  There is correspondingly the
probability $p$ that the atom is found to the left, and $1-p$ that
the atom is round to the right, of the barrier. \item The barrier
is then isothermally, reversibly moved to the centre of the box,
requiring a mean work expenditure of
\[
E^\prime = (1-H) k T \ln 2
\]
\end{itemize}
Now the final statistical state of the one-atom gas after $RLE(p)$
is identical to the statistical state before $LE(p)$.  The net
work required for $LE(p)$ exactly matches the net work extracted
by $RLE(p)$ and the entire process has been achieved in a
thermodynamically reversible manner.

It is important to note that the $LE(p)$ operation is not
conditional upon the location of the atom.  A measurement of, or
correlation to, the logical state of the atom is not required.  It
is only tailored to the value of $p$, to the amount of information
that is being reset. If the value of $p=0$ or $p=1$ the net work
required for $LE(p)$ or $RLE(p)$ is zero. These correspond to the
cases where the bit is known to be in a particular state, so the
Shannon information is zero.  In these cases, it is well known
that operations exist ("Do nothing" and "NOT", for example) that
can reset the bit to zero non-dissipatively.

The tailoring of the $LE(p)$ operation to the specific amount of
information being reset makes it difficult to use this in
practice. It is unlikely that an engineer designing a computer
will be able to determine the statistical probabilities of the
inputs to the logical operations the computer will perform. Simply
assuming a uniform probability in all cases will not work. If, for
example, the input to one operation was itself the output of an
AND operation, then one or the other operation must have
non-uniform probabilities for its input states.

Nevertheless, the key conclusion remains: in principle it is
always possible to perform a logically irreversible operation in a
thermodynamically reversible manner.  Clearly we have not
explicitly shown that {\em all} logically irreversible operations
can be implemented as thermodynamically reversible processes, but
provided that the thermodynamic process is tailored to the amount
of information that is being manipulated, we see no reason that
this cannot be done for any logically irreversible operation.

Perhaps the correct generalisation of Landauer's principle should
be:
\begin{quote}
A logical operation needs to generate heat equal to at
least\footnote{If the change in Shannon information is negative,
then this gives the {\em maximum } amount of heat energy that can
be {\em extracted as work } during the process.} $k T \ln 2$ times
the change in the total quantity of Shannon information over the
operation.
\end{quote}

Neither logical reversibility nor erasure of information appears
in this formulation.  {\em Any } operation may, in principle, be
implemented as a physical process which reaches the limit and so
can be made thermodynamically reversible.
\section{Logical Computation}\label{computation}

In the previous Section we considered the thermodynamic reversal
of $LE$ and its generalisation to non-uniform probability
distributions.  We found that this could always be made
thermodynamically reversible.  We will now consider these as
logical operations (for simplicity we will use $p=\frac{1}{2}$
from now on, although the conclusions can easily be generalised to
the non-uniform case).

\begin{table}
\centering
    \begin{tabular}{c||c}
        \multicolumn{2}{c}{LE} \\ \hline
        IN & OUT \\
        \hline \hline 0 & 0 \\ 1 & 0 \\
        \multicolumn{2}{c}{} \\
        \multicolumn{2}{c}{} \\
    \end{tabular}
    \begin{tabular}{c||c}
        \multicolumn{2}{c} {RLE} \\ \hline
        IN & OUT \\
        \hline \hline 0 & 0 \\ 0 & 1 \\
        \multicolumn{2}{c}{} \\
        \multicolumn{2}{c}{} \\
    \end{tabular}
    \begin{tabular}{c||c}
        \multicolumn{2}{c}{RAND}\\ \hline
        IN & OUT \\
        \hline \hline 0 & 0 \\ 0 & 1 \\ 1 & 0 \\ 1 & 1
    \end{tabular}
    \caption{Logical Operations \label{tb:logic}}
\end{table}

Logical reversibility is defined by Landauer (1961):
\begin{quote}
We shall call a device logically irreversible if the output of a
device does not uniquely define the inputs.
\end{quote}
It is clear from the truth tables (Table \ref{tb:logic}) that the
operation $LE$ is logically irreversible. $RLE$, on the other
hand, {\em is} logically reversible, if rather trivially so (given
the output one can define what the input must have been - it must
have been zero!)

If we now combine the two operations $(RLE)(LE)$ we get a new
logical operation $RAND$.  This is also logically irreversible.
The logical state of the bit after the computation cannot be used
to deduce the logical state of the bit before the computation.
Whatever information was originally represented by the bit has
been lost. An `erasure' of information has taken place (it must
have taken place as Landauer {\em Erasure} is the first part of
the process.  Information that is `erased' cannot be recovered!).

However, during $(LE)$ $kT \ln 2$ heat is generated, while during
$(RLE)$ $kT \ln 2$ is {\em extracted} from the environment.  The
final thermodynamic state is identical to the initial
thermodynamic state\footnote{It is accepted by both Bennett (2003)
and Leff and Rex (1990) that $LE(\frac{1}{2})$ is
thermodynamically reversible. We have here simply reversed the
thermodynamic process.}. So no net generation of heat has taken
place. Nevertheless, a logically irreversible operation, involving
the loss of 1 bit of information, has clearly taken place.

This directly contradicts the most common manner in which the
generalisation of Landauer's principle is stated:
\begin{quote}
To erase a bit of information in an environment at temperature $T$
requires dissipation of energy $\geq kT \ln 2$. (Caves 1990, 1993)

in erasing one bit \ldots of information one dissipates, on
average, at least $k_B T \ln\left(2\right)$ of energy into the
environment. (Piechocinska, 2000)

a logically irreversible operation must be implemented by a
physically irreversible device, which dissipates heat into the
environment (Bub, 2002)
\end{quote}
We have shown that logically irreversible operations do not {\em
always} imply a dissipation of energy into the environment.

In case it is argued that there is still {\em some} heat
generation - during the $LE$ operation - but that we have just
recovered this at a later stage, we should note that the combined
operation $RAND$ can be achieved without any work being required
at any stage:

\begin{itemize}
\item Remove the barrier from the centre of the box. \item Wait
for a time larger than the thermal relaxation time of the atom.
\item Insert the barrier back into the centre of the box.
\end{itemize}

Now we have created exactly the same truth table as $(RLE)(LE)$ -
so the original bit of information must again be considered lost
or erased - and the statistical state remains the same at the end
as at the beginning of the operation.  The operation is logically
irreversible, but is thermodynamically reversible and undergoes no
heat exchange with the environment.

What we are seeing here is the consequences of one of the two
facts we noted about Landauer Erasure above: the erasure stage of
the operation is not the heat generation stage. It is the
resetting operation (which compresses the probability
distribution) which generates heat.  The erasure stage, in which
the logical irreversibility actually takes place and the original
information is lost, does not require an exchange of energy with
the environment.

\pict{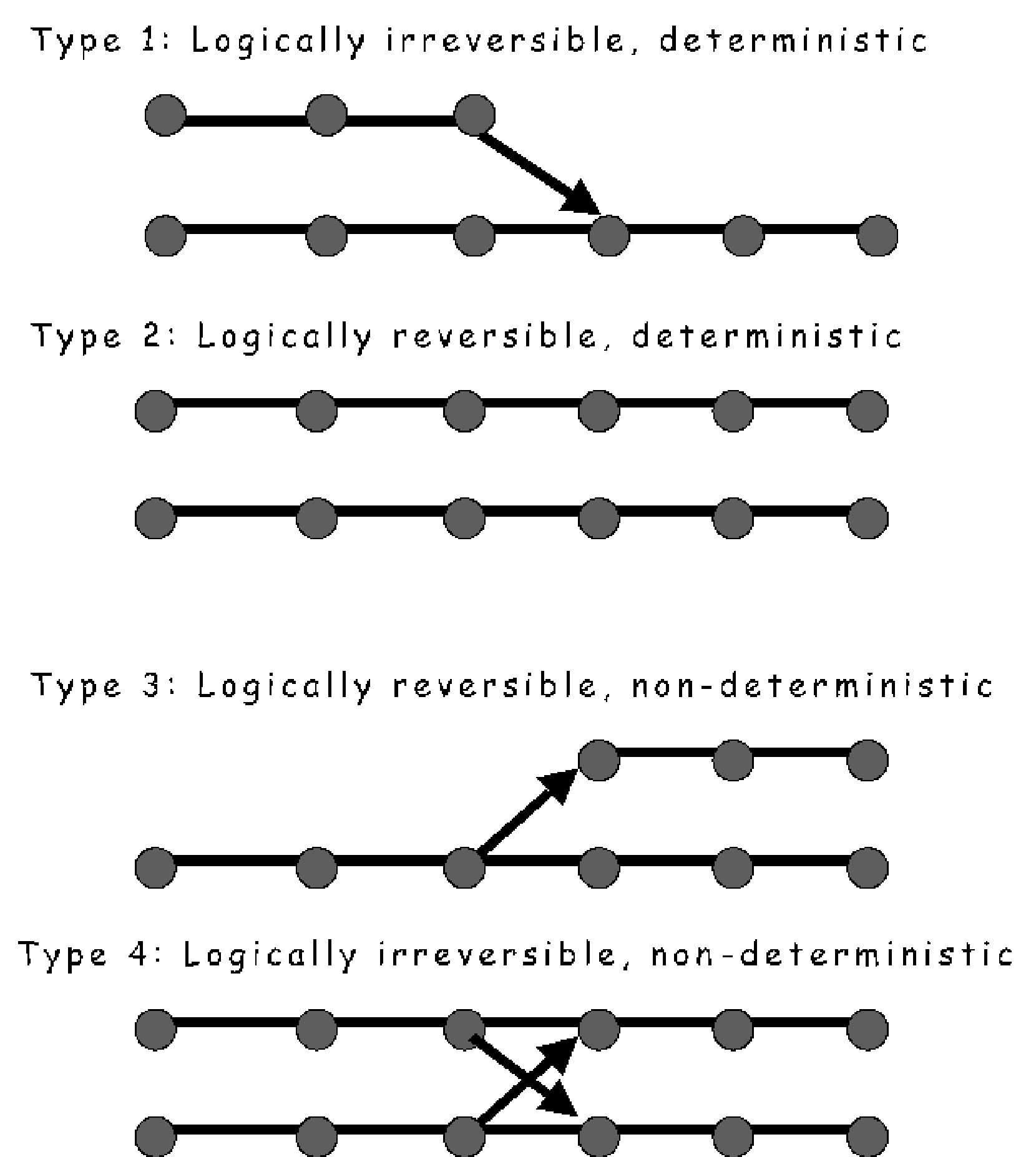}{Computational Paths}

\subsection{Non-deterministic operations}

The operations $RLE$ and $RAND$ have unusual "truth tables". They
certainly do not appear in most works on computational logic! The
reason for this is that they display the converse property to
logical irreversibility: logical non-determinism\footnote{It is
noticeable that in Bennett (2003), he refers throughout to {\em
deterministic} computation.}. We will define this as follows:

\begin{quote}
We shall call a device logically non-deterministic if the input of
a device does not uniquely define the outputs.
\end{quote}

This should not be taken as requiring some kind of fundamental
physical process that is non-deterministic (such as quantum
wavefunction collapse) any more than logical irreversibility
requires a fundamental physical process that is irreversible.  In
both cases an interaction with an uncorrelated environment (such
as a heat bath) is all that is required.

\subsection{Computational Paths}
Let us consider the computational paths that can be constructed
from the addition of non-deterministic logical operations (Figure
\ref{fg:figure3}).

\subsubsection{Irreversible, Deterministic}
Type 1 computational paths are the most commonly encountered. In
this we consider two possible input states (which for convenience
we will assume are equiprobable) and at some point a merging of
the computational paths takes place (such as by $LE$). This
requires a minimum amount of work to be performed on the system,
which is expelled as heat.
\subsubsection{Reversible, Deterministic}
Type 2 paths are, in some respects, an improvement.  Using
reversible logic gates, such as the Fredkin-Toffoli gate, the
output from the logical operation will contain additional bits
which can be used to reconstruct the input states.  Logical
irreversibility is eliminated and no heat need be generated.  All
Type 1 computations may also be performed by Type 2 computational
paths, as shown by Bennett (1973).

Type 1 and 2 computational paths are deterministic.  If one
restricts ones consideration to deterministic logical operations,
then logical irreversibility implies heat generation. With the
addition of non-deterministic logical operations, we find new
computational paths are possible, whose thermodynamic properties
are different.

Should we confine ourselves to deterministic logical operations?
It seems unnecessarily restrictive to do so.  Aside from the
simple fact that such operations exist (and may even have
computational value) there is the fact that the non-deterministic
operation $RLE$ is just the thermodynamic reverse of $LE$. Leff
and Rex (1990) and Bennett (2003) accept that $LE$ is
thermodynamically reversible. For this to be meaningful, one must
accept the reverse thermodynamic process as a legitimate
operation.  This reverse process is $RLE$ and is
non-deterministic.

\subsubsection{Reversible, Non-deterministic}
Type 3 paths are logically reversible but non-deterministic.  An
example of such a process is the reverse Landauer Erasure,
$RLE(p)$.  It is interesting to note that this operation can
extract heat from the environment and store it in a work
reservoir.  It cools a computer down, in contrast to Type 1
operations which heat computers up.  Type 3 paths are generally
the reverse of Type 1 paths (and vice versa).

This produces an interesting result.  In Bennett (1973) a general
procedure is devised for simulating Type 1 computations by Type 2
computations and so avoiding the generation of heat. This involves
copying the output of the initial simulation\footnote{This applies
only to {\em classical} computation, see Maroney (2001,2004).},
then reversing all the logical steps, leaving only the input state
and the copy of the desired output.

It can now be seen that there is an equivalent way of running a
computer at a minimal thermodynamic cost, but without the
restriction to using logically reversible computation. Perform the
Type 1 computation, using irreversible logical operations and
expelling heat into the environment. Then copy the output as
before and perform the Type 3 computation that is the exact
reverse of the original Type 1 computation.  Now all the heat
generated during the Type 1 computation can be extracted to
perform work during the Type 3 computation!

This differs from the reversible simulation using Type 2 paths in
two ways:
\begin{itemize}
\item For reversible simulation, the final state is the desired
output and the original input state.  With the non-deterministic
reversal, the final state is the desired output plus a random
choice of one of the possible input states that could produce the
given output state, with a probability given by the statistical
distribution of the input states. \item To achieve the minimal
thermodynamic cost, it is necessary for the implementation of the
separate irreversible and non-deterministic operations to be
tailored to the statistical distribution of the input states.
This would be very hard to achieve in practice.
\end{itemize}
Bennett's procedure for simulating Type 1 computations by Type 2
computations is therefore a far superior method of eliminating the
conversion of work into heat than using a Type 3 computation to
extract the generated heat back again.
\subsubsection{Irreversible, Non-deterministic}
Type 4 computational paths are irreversible and non-deterministic.
Such computations may sometimes require a generation of heat, may
sometimes be able to extract energy from the environment and may
sometimes be implemented with no exchange of energy, depending
upon the exact nature of the manipulation of the information.  In
the case of the $RAND$ operation, no exchange of energy is
required.

There is a curious asymmetry between irreversibility and
non-determinism.  Type 1 calculations can be performed by Type 2
operations. Type 3 and 4 operations, however, cannot be performed
by Type 1 or 2 operations.  This is because the non-deterministic
operation can effectively operate as a random number
generator\footnote{Let it be clear the source of the random number
is not necessarily through a non-determinism in the fundamental
laws of physics.  It is a computational non-determinism whose
source is sampling the uncorrelated degrees of freedom of the
environment.}.  A Type 3 or 4 operation can increase the
algorithmic complexity of an input bit string.  This is not
possible for Type 1 or 2 operations.

This is especially curious when we consider that a Type 3
operation is just the reverse of a Type 1 operation, but if we
reverse the simulation of a Type 1 operation by a Type 2
operation, we do {\em not} get a Type 3 operation!

All these computational paths can be accomplished in a
thermodynamically reversible manner.  However, one cannot tell
directly from the truth table alone what the reversible
thermodynamic implementation will be - that requires the
statistical distribution of the input states.  The exception to
this is Type 2 computations, which require neither generation nor
extraction of energy, regardless of the statistical distribution.
\section{Shannon Information}\label{shannon}
We will now briefly consider how the difference between Type 2 and
Type 4 relates to the processing of Shannon information.

Start with the usual source and signal states occurring with
probability $p$.  Now suppose that this is arranged so that the
energy of the states happens to be $E_0+kT \ln p$ where $T$ is the
temperature of the environment and $E_0$ is some constant.  Now
further suppose that the communication channel is very noisy. In
fact over the course of the signal propagation, any individual
state becomes completely thermalised.

The statistical state at the end of the transmission will be
identical to the statistical state at the start of the
transmission. However, one cannot say that the receiver is
actually gaining any information about the signal {\em
transmitted}.  For information transmission, the individual state
needs to be made stable against noise.  Otherwise, it is not a
signal at all.

For the thermodynamic state, there is no change.  The
thermodynamic entropy is the same at the start as at the end. Such
thermalisation through contact with the environment is often an
essential part of thermodynamic processes (such as isothermal
compression and expansion of a gas). The individual state has no
significance.

Of course, it remains the case that the receivers ignorance as to
which state will be decoded remains $-\sum p \ln p$.  What has
changed is the significance, if any, the receiver may attach to
the signal.

Consider a Type 2 computational paths and a Type 4 computational
paths, where they differ only in the existence of a single $RAND$
operation in the Type 4 path.  Both before and after the $RAND$
operation the statistical state of the Type 4 path is the same as
the statistical state of the Type 2 path.

The quantity of Shannon information has not been changed by the
$RAND$ operation.  This cannot be taken to mean that Type 2 and
Type 4 paths are informationally equivalent!  From the point of
view of information processing, they are fundamentally very
different types of operation.  From the point of view of
thermodynamics, however, the $RAND$ operation has had no effect.
\section{The Szilard Engine}\label{szilard}
Now let us turn to the Szilard Engine, and the use of Landauer's
principle in the exorcism of Maxwell's demon (Figure
\ref{fg:figure4}). A single atom is contained within a box, with a
moveable barrier (or piston) inserted in the centre of the box.
There is a probability $\frac{1}{2}$ of the atom being on either
side of the barrier.  The demon has a single memory register,
which is initially in logical state $0$.

The normal presentation of this, in terms of information, might go
as follows:

\begin{enumerate}
\item The Demon performs a measurement upon the box to determine
the location of the atom.  The measurement generates no heat. The
state of the Demon's memory now represents the location of the
atom.  As information has been gained about the location of the
atom, its entropy has been reduced. \item Having determined the
location of the atom, the piston is moved to one side or the
other, and the isothermal expansion used to lift a weight.  This
extracts $kT \ln 2$ energy, corresponding to the reduction in
entropy of the atom during the measurement. \item At first it
seems that the 2nd Law has been violated.  However, there is still
the information about the location of the atom recorded in the
Demon's memory.  To complete the cycle and return to its initial
state, this record must be erased. According to Landauer's
principle this requires the conversion of $k T \ln 2$ work into
heat, exactly compensating for the work extracted.
\end{enumerate}
\pict{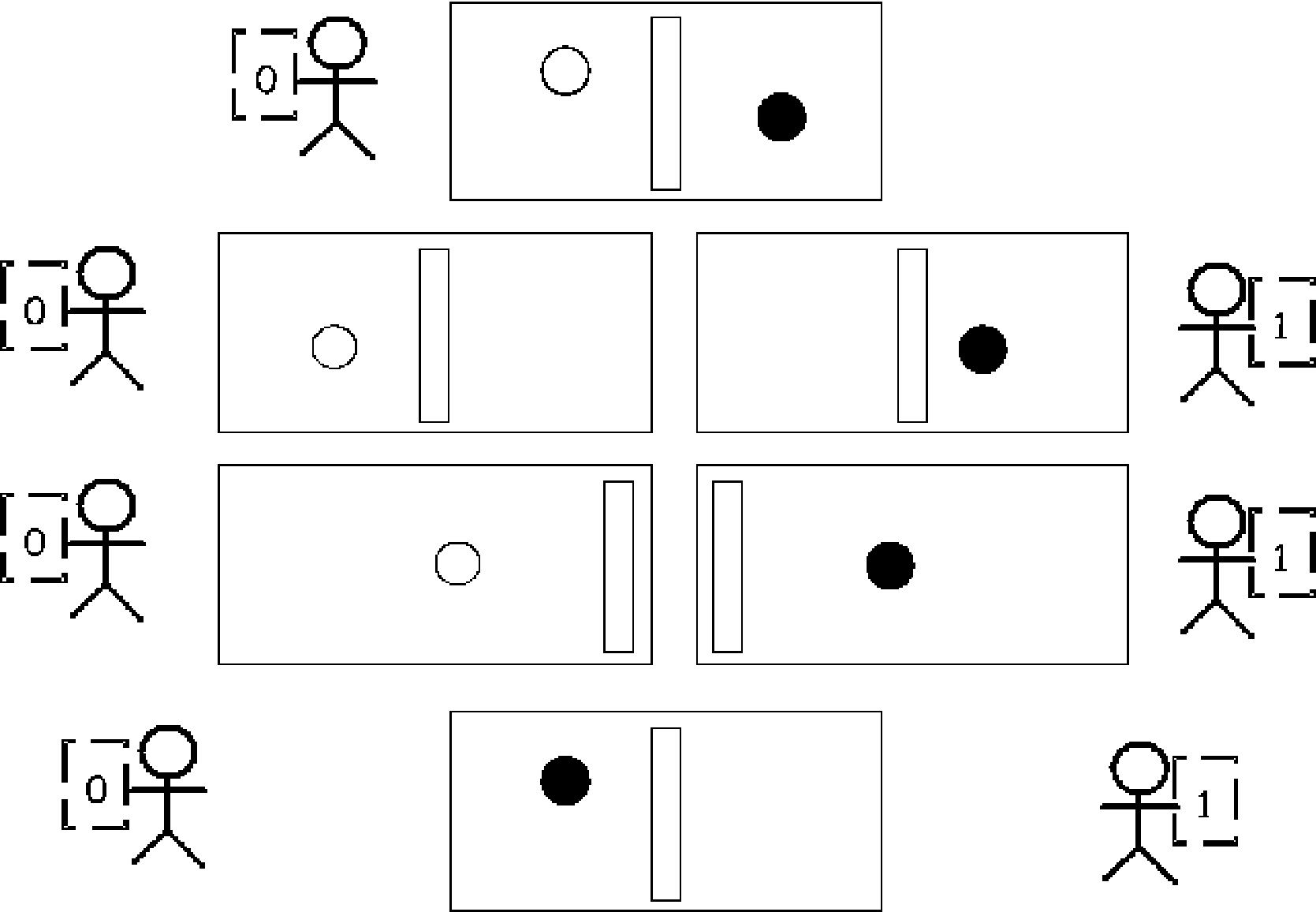}{Landauer Erasure and the Szilard Engine}

Now let us consider how this is affected by the arguments of
previous sections.

As we have seen, the Demon's one-bit memory can itself be
physically realised by an atom in a box with a moveable barrier.
Instead of resetting the Demon's memory, by an $LE$ operation, let
us perform the $RAND$ operation upon the Demon's memory, by
pulling out and reinserting the barrier in the Demon's memory
register. The result is a complete randomisation of the state of
the Demon (Figure \ref{fg:figure5}).

This operation is thermodynamically reversible and leaves the
statistical state unchanged. Indeed, from the point of view of
thermodynamics, nothing has happened. If there was a problem for
thermodynamics, requiring resolution, before the $RAND$ operation,
that problem still exists after the $RAND$ operation.

\pict{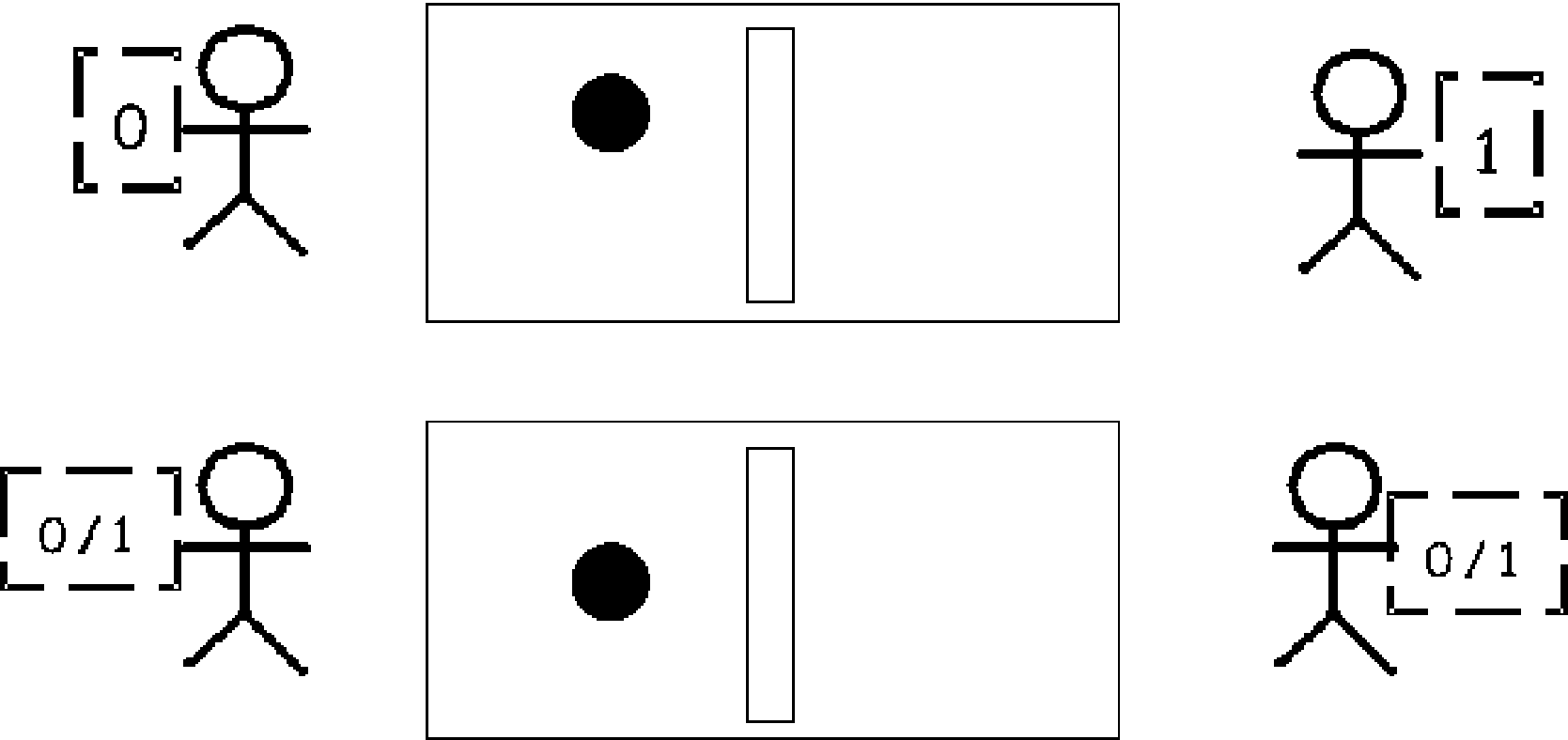}{Randomisation of Demons "Memory"}

However, it causes problems for the information erasure story.
There is no correlation between the state of the Demon's memory
after the $RAND$ and the state of the memory (or the initial
location of the atom) before $RAND$.  The one-bit "memory" in the
demon no longer represents the location of the atom.  In fact it
no longer represents information about anything.  The $RAND$
operation has already `erased' the record of the Demon's
measurement, it has `erased' the information about the location of
the atom, but has not generated any heat.

Is this a problem for thermodynamics?  Hardly.  The energy $kT \ln
2$ has been extracted, it is true.  What is required, to `save'
the second law, is that there be a compensating increase of
entropy of $k \ln 2$, and this has already happened.  At the start
of the process the Demon's `memory' was in the logical state $0$.
At the end of the process it may be in either logical state with
equal probability.  Its entropy has increased by $k \ln 2$.  This
is precisely what is needed.  As the Szilard Box itself is now
back in its initial state, the entire process could simply be
regarded as convoluted procedure for extracting $k T \ln 2$ energy
from the isothermal expansion of the {\em Demon's} state, by an
operation such as $RLE$!

If it is not necessary to refer to Landauer Erasure after the
$RAND$ operation, then it was not necessary to refer to it
beforehand.  Why then has Landauer Erasure been promoted so
strongly as the necessary principle to resolve the Szilard Engine?
Historically, it was suggested by Brillouin (1951) and Gabor
(1964) that the act of information acquisition (or performing a
measurement) by the Demon must dissipate $kT \ln 2$ energy into
the environment. It was Landauer's 1961 paper that demonstrated
the falsity of this argument - they had wrongly generalised from
certain measurement processes, which were dissipative, to assume
all measurement processes must generate heat.

Having eliminated the generation of heat during the measurement
process, it is natural to look for a different source of heat
generation, and in the same paper Landauer appears to provide the
answer: resetting is a heat generating operation, and requires the
conversion of exactly the right amount of work into heat.

\begin{quote}
The essential irreversible act, which prevents the demon from
violating the second law, is not the measurement itself but rather
the subsequent restoration of the measuring apparatus to a
standard state (Bennett, 1982)

information processing and acquisition have no intrinsic,
irreducible thermodynamic cost, whereas the seemingly humble act
of information destruction does have a cost, exactly sufficient to
save the Second Law from the Demon. (Bennett, 2003)
\end{quote}

This misses the point that it is not actually necessary to
generate heat (or perform an irreversible act) to `save the Second
Law', it is only necessary to have a compensating {\em entropy}
increase somewhere. The compensating increase in entropy has
already taken place, and has taken place in the Demon itself.

If the increased entropy of the Demon is sufficient to save the
second law after the $RAND$ operation, when the memory register no
longer represents information about the atom, then the fact that
the memory register happens to represent information about the
atom before the $RAND$ operation is irrelevant to the resolution
of the problem.  The appeal to information acquisition, memory and
erasure obscures the real reason why the second law is not
violated. The movement of the barrier has to be correlated to the
location of the atom.  To do this requires an auxiliary system.
This auxiliary ends up in a higher entropy state.  This is not a
principle of computation or of logic - it is a consequence of the
{\em physics} of Hamiltonian dynamics.  It is immaterial whether
one considers the auxiliary to be a `memory cell' or whether it
can be regarded as representing `information' about the state of
the atom, or anything else.
\section{Conclusion}\label{conclusion}
Logically reversible operations may always be physically
implemented as a thermodynamically reversible process. They may
also, by being sub-optimally designed, be physically implemented
as thermodynamically irreversible processes.

We have shown the same holds true for logically irreversible
operations.  Any given quantity of information can be reset to
zero in a thermodynamically reversible manner.  For the limit of
$kT \ln 2$ heat generation per bit to be reached, the
thermodynamic process {\em must} be reversible.

In practice, logical operations are implemented by sub-optimal
physical processes and so are thermodynamically irreversible.
However, this irreversibility is not caused by the nature of the
logical operation, it is by way of the operation being implemented
by a thermodynamically sub-optimal physical process. This is as
true for logically irreversible operations as it is for logically
reversible operations.

This does not contradict Landauer (1961) in the least.  All that
Landauer can be said to have shown was that a resetting operation
required a generation of heat in the environment. However, a
confusion then appears to arise through the incorrect use of the
term `dissipation'.  In Landauer (1961) and in much of the
surrounding literature `dissipation' is used more or less
interchangeably with `heat generation'.  Strictly, dissipation
should be used only when the conversion of work to heat arises
through dissipative forces (such as those involving friction)
which are thermodynamically irreversible. Forces which are
thermodynamically reversible are non-dissipative\footnote{The
author would like to thank John Hannay for discussions on this
point.}.

As an example, consider the isothermal compression of an N-atom
ideal gas to half it's volume.  If this takes place sufficiently
slowly then work equal to $N k T \ln 2$ must be performed upon the
gas.  The internal energy of the gas remains constant at
$\frac{3N}{2}k T$ throughout and the work done is converted to
heat in the heat bath.  All this heat generation is
thermodynamically reversible as the free energy of the gas
increases by $N k T \ln 2$ and the net change in entropy of the
gas and heat bath is zero. In this case it would be incorrect to
refer to the conversion of $N k T \ln 2$ work into heat as
`dissipation', as the force is frictionless, thermodynamically
reversible, and so `non-dissipative'.

In the case where $N=1$ we have exactly the isothermal compression
of a one-atom gas that is used in Landauer Erasure.  It should be
clear, therefore, that when Landauer's minimum limit of $k T \ln
2$ heat generation is reached this is {\em non-}dissipative.  This
can make statements such as
\begin{quote}
To erase a bit of information in an environment at temperature $T$
requires dissipation of energy $\geq kT \ln 2$ (Caves 1990,1993)
\end{quote}
hard to understand when the equality is reached (and all the
literature agrees that the equality can, in principle, be
reached).

This confusion over the use of the term `dissipation' seems to be
the basis of Shenker's (2000) criticism of Landauer Erasure (which
Shenker refers to as the `Landauer Dissipation Thesis'), as
Landauer Erasure can always be made `non-dissipative'. In contrast
in this paper we have avoided the use of the term `dissipation'
where possible, and used instead `heat generation'. Where
`dissipation' is used in the literature, we have assumed that
`heat generation' is all that was intended.  Certainly Landauer
(1961) can consistently be understood this way.

Nevertheless, the use of this term seems to have created a faulty
link from `heat generating' to `thermodynamically irreversible':
\begin{quote}
The existence of logically reversible automata suggests that
physical computers might be made thermodynamically reversible, and
hence capable of dissipating an arbitrarily small amount of energy
(Bennett, 1973)

a logically irreversible operation must be implemented by a
physically irreversible device, which dissipates heat into the
environment (Bub, 2002)
\end{quote}

We have clearly shown in this paper that logically irreversible
devices are as capable of being made thermodynamically reversible
(and hence non-dissipative) as logically reversible devices.

There is then a second confusion regarding whether logically
irreversible operations must {\em all} be heat generating.  What
Landauer (1961) argues is that all {\em unavoidably} heat
generating operations are logically irreversible.  This is quite
correct, but this does not mean that all logically irreversible
operations are unavoidably heat generating.  As we have seen, {\em
some} logically irreversible operations need not generate heat.
Some may even absorb heat and convert it to useful work!

We have seen that these operations remove any need to consider
Landauer erasure or concepts of information processing, when
resolving the `problem' of the Szilard Engine.  We would also
argue that this demonstrates a conceptual difference between
information and entropy.

The measure of the Shannon information clearly has the same
mathematical form as the measure of the Gibbs entropy.  Sharing
the same mathematical form, however, does not make two things the
same.  An electrical circuit composed of a capacitor and an
inductor can be described by the same mathematical equations as a
mass on a spring, but this does not make mass the same thing as
inductance.

The logically irreversible, non-deterministic $RAND$ operation has
a profound affect upon the informational state of a system, while
leaving its thermodynamic state unaffected.  The fact that the
{\em quantity} of the Shannon information is the same before and
after $RAND$ should not lead us to think nothing has happened:
Type 4 computational paths are profoundly different from Type 2
computational paths, as information processing systems.  For
thermodynamics, however, there is no difference.  So there is more
to a concept than its measure: information and entropy are not the
same thing.

%\bibliographystyle{alpha}
%\bibliographystyle{apacite}
%\bibliography{paper8}

\end{document}